\documentclass[reprint, amsmath,amssymb, prl,floatfix,superscriptaddress]{revtex4-2}

\usepackage{graphicx}
\usepackage{bm}
\usepackage{mathrsfs}  
\usepackage[usenames,dvipsnames]{xcolor}
\usepackage[colorlinks=true,citecolor=blue]{hyperref} 

\newcommand{\figref}[1]{Fig.\,\ref{#1}}

\makeatletter
\DeclareRobustCommand{\element}[1]{\@element#1\@nil}
\def\@element#1#2\@nil{%
  #1%
  \if\relax#2\relax\else\MakeLowercase{#2}\fi}
\makeatother

\begin{document}

\title{Atomic-scale probe of molecular magneto-electric coupling}

\author{Mohammad Amini}
\affiliation{Department of Applied Physics, Aalto University, FI-00076 Aalto, Finland}

\author{Linghao Yan}
\affiliation{State Key Laboratory of Bioinspired Interfacial Materials Science, Institute of Functional Nano \& Soft Materials (FUNSOM), Soochow University, 215123 Suzhou, China}
\affiliation{Department of Applied Physics, Aalto University, FI-00076 Aalto, Finland}

\author{Orlando J. Silveira}
\affiliation{Department of Applied Physics, Aalto University, FI-00076 Aalto, Finland}

\author{Adolfo O. Fumega}%
\affiliation{Department of Applied Physics, Aalto University, FI-00076 Aalto, Finland}

\author{Viliam Va\v{n}o}
\affiliation{Department of Applied Physics, Aalto University, FI-00076 Aalto, Finland}

\author{Jose L. Lado}%
\affiliation{Department of Applied Physics, Aalto University, FI-00076 Aalto, Finland}

\author{Shawulienu Kezilebieke}
\affiliation{Department of Physics, Department of Chemistry and Nanoscience Center, 
University of Jyväskyl\"a, FI-40014 University of Jyväskyl\"a, Finland}

\author{Peter Liljeroth}
\email{Corresponding authors. Email: peter.liljeroth@aalto.fi, robert.drost@aalto.fi}
\affiliation{Department of Applied Physics, Aalto University, FI-00076 Aalto, Finland}

\author{Robert Drost}
\email{Corresponding authors. Email: peter.liljeroth@aalto.fi, robert.drost@aalto.fi}
\affiliation{Department of Applied Physics, Aalto University, FI-00076 Aalto, Finland}

\date{\today}

\begin{abstract}
Van der Waals heterostructures are a core tool in quantum material design. The recent addition of monolayer ferroelectrics expands the possibilities of designer materials. Ferroelectric domains can be manipulated using electric fields, thus opening a route for external control over material properties. In this paper we explore the possibility of engineering magneto-electric coupling in ferroelectric heterostructures by studying the interface of bilayer SnTe with iron phthalocyanine molecules as a model system. The molecules act as sensor spins, allowing us to sample the magneto-electric coupling with nanometer precision through scanning tunneling microscopy. Our measurements uncover a structural, and therefore material-independent and intrinsic, mechanism to couple electric and magnetic degrees of freedom at the nanoscale.
\end{abstract}
\maketitle

The family of two-dimensional materials now has members from most electronic phases of matter, including superconductors, correlated materials, and magnets. Due to this wide variety of states, heterostructures of two-dimensional materials, with their tunable properties and large contact surfaces, have become an important tool in quantum material design \cite{novoselov20162d, liu2016van, cao2018unconventional, kezilebieke2020topological, vavno2021artificial}. The recent discovery of two-dimensional ferroelectrics brings new opportunities in the field of quantum material design \cite{Chang2016, yasuda2021stacking, vizner2021interfacial, Amini2023}. 

As with bulk materials, two-dimensional ferroelectrics develop a spontaneous electric polarisation below a critical transition temperature \cite{Hook}. The ferroelectric ground state breaks the original lattice symmetry and leads to the formation of ferroelectric domains, in analogy to ferromagnetism. The ferroelectric domains can be manipulated using electric fields \cite{Amini2023, Chang2016}, opening a route for external control in a device setting. Interestingly, the critical temperature of the prototypical ferroelectric transition in SnTe in the monolayer limit is ca. 270\,K, making it extremely attractive from an applications point of view \cite{Chang2016}.

Adjacent ferroelectric domains will generally have different stacking with the underlying substrate. In few layer thick samples, this leads to further strain from the different adsorption geometries, which result in variations of the crystal field. These variations offer a mechanism to engineer magneto-electric coupling in two-dimensional materials by exploiting local changes in the magneto-crystalline anisotropy (MCA). Such magneto-electric coupling provides a quick and convenient route to control magnetic properties through electric fields\cite{Fiebig2016, tang2025towards}. From a more fundamental perspective, the competition between several order parameters brings new possibilities to designing novel quantum states of matter \cite{sachdev2000quantum, Fiebig2016, matsubara2015}.

MCA plays a large role in shaping magnetism at the nanoscale \cite{Ternes2017, Ternes2015}. In the presence of a crystal lattice, spin systems are more readily polarized along certain preferential spatial directions. These often coincide with the high-symmetry axes of the host crystal. Spin systems can be highly sensitive even to small changes in the crystal field \cite{Ternes2017, bryant2013local, heinrich2015tuning, cheng2021situ}. Exploiting local changes in the MCA is a simple and promising route to control magneto-electric coupling in any material, and exceptionally suitable to leverage from two-dimensional materials. 

\begin{figure}[!h]
    \centering
    \includegraphics[width =1 \columnwidth]{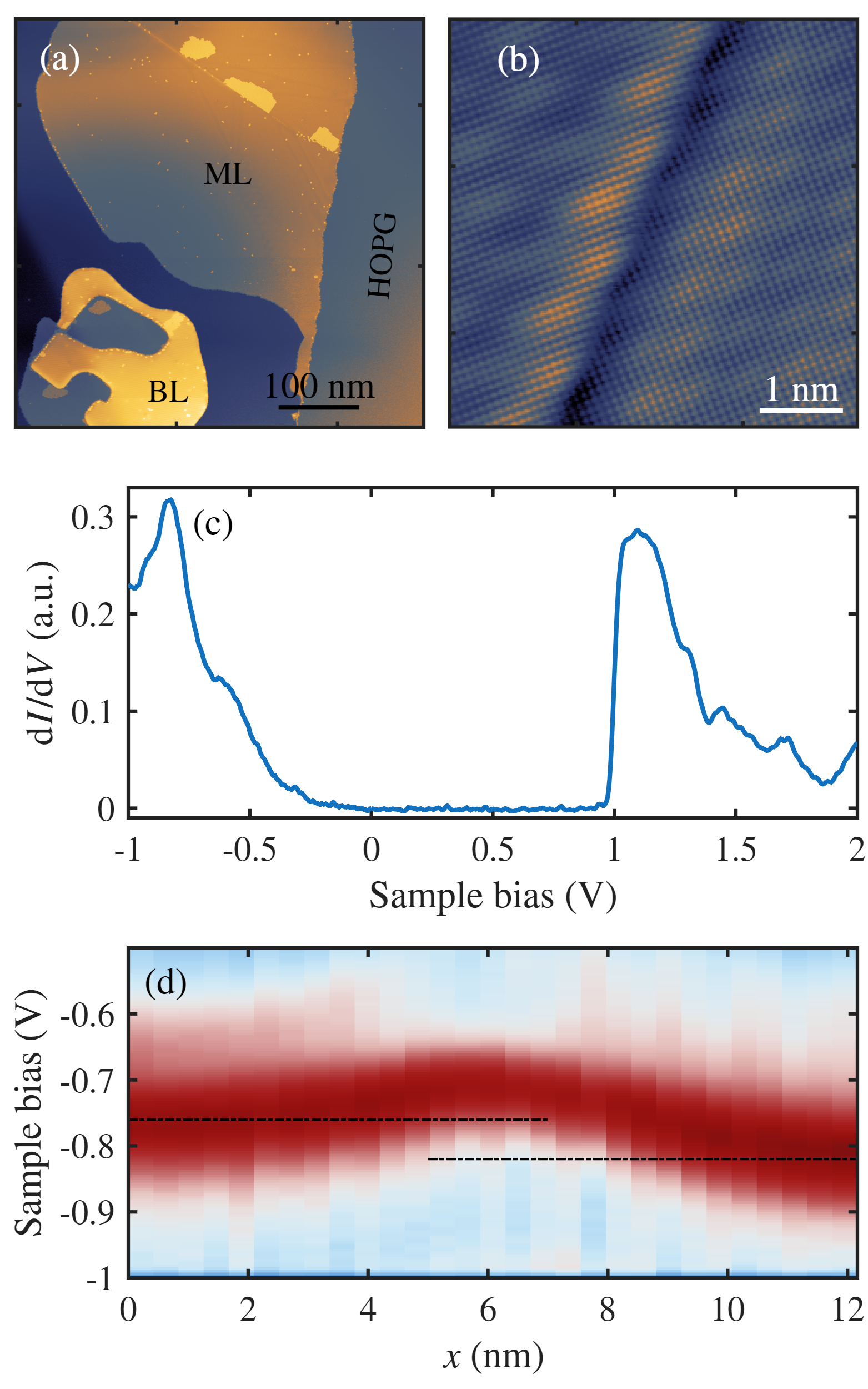}
    \caption{(a) Typical sample with ML and BL islands of SnTe on HOPG. Agglomerations of FePc decorate the SnTe (1\,V, 12\,pA). (b) Example of a ferroelectric grain boundary in BL SnTe. The 90$^{\circ}$ degree lattice rotation leads to stark differences in the observed moir{\'e} patterns in the adjacent domains (1\,V, 24\,pA). (c) Representative conductance spectrum acquired on BL SnTe. The conduction band onset lies at ca +1 V, the valence band onset at ca. -0.23 V. The valence band maximum (VBM) is found at much lower energies, ca. -0.8 V.  (d) Waterfall plot tracking the energy of the VBM across a grain boundary. Owing to variations in lattice strain, the VBM settles to different values in either domain.} 
    \label{fig:Figure1}
\end{figure}

In this paper we demonstrate an engineered magneto-electric coupling at the interface of bilayer (BL) SnTe with iron phthalocyanine (FePc) molecules as a model system. The FePcs act as sensor spins, allowing us to sample the magneto-electric coupling with nanometer precision through scanning tunneling microscopy (STM) \cite{wang2021symmetry, zhang2022electron}. We find clear correlations between the ferroelectric domain structure of SnTe and the spin excitation energies of FePc. Our experiments deliver a proof of concept for a new class of engineered two-dimensional heterogeneous multiferroics.

We grew SnTe samples by direct depostion of SnTe in UHV (see methods section for details). \figref{fig:Figure1}a shows a typical sample with islands of SnTe rising as mesas from the HOPG substrate. Monolayer (ML) and BL areas coexist on the sample. We readily observe the ferroelectric domain walls typical of SnTe, see \figref{fig:Figure1}b. Conductance spectroscopy further confirms the semiconducting nature of SnTe, with a representative spectrum presented in \figref{fig:Figure1}c. The conduction band has a sharp onset at 1\,V, with the valence band appearing more gradually beginning from -0.5\,V. The band gap, $\sim 1.5$\,eV, is somewhat reduced from the bulk value.

The most relevant feature of SnTe in the context of this paper is the spontaneous formation of ferroelectric domains at low temperatures \cite{Chang2016, Amini2023}. The material undergoes a structural phase transition whereby the Sn and Te sub-lattices are displaced with respect to each other. The critical temperature depends on the thickness of the material and is considerably increased in the few layer limit. The distortion breaks the cubic lattice symmetry of SnTe and this reduction in symmetry, in turn, leads to the formation of non-equivalent ferroelectric domains in each island. The lattice distortion is directly evident through measurements of the lattice constants from STM topographs (see Supporting Information (SI) for details). 

The prominent domain walls  (see \figref{fig:Figure1}b) further highlight the ferroelectric ground state. These domain walls appear as lines of brighter or darker contrast in topographic images, depending on the applied bias voltage. This contrast arises from band bending near the domain boundary \cite{Chang2016, Amini2023}.

There is another, more subtle, contrast between adjacent domains. These do not appear at the same apparent height in STM scans, despite being nominally identical except for a rotation of the polarization axis by 90\,$^{\circ}$. These observations are a first indication that adjacent ferroelectric domains are in fact not equivalent. A closer look at the electronic structure through tunnelling spectroscopy confirms this. \figref{fig:Figure1}d shows a series of line spectra taken perpendicular to a domain wall as a waterfall plot. As expected, the conduction band edge rises to higher energies in the vicinity of the domain wall. It does not, however, settle to identical values deep in the domains on either side. 

We hypothesize that interactions with the HOPG substrate are the driving force behind these local changes. As is typical for two-dimensional materials, thin layers of SnTe form a moir\'{e} pattern with the substrate \cite{andrei2021marvels, kennes2021moire}. Since the two lattices belong to different symmetry classes, the effective lattice rotation of 90\,$^{\circ}$ between adjacent domains leads to pronounced changes in the stacking order between substrate and adlayer changes. The moiré pattern formed by SnTe and HOPG amplifies these difference. Indeed, we observe vast changes in the period and angle of the moiré patterns in adjacent domains, see \figref{fig:Figure1}b.  The ferroelectric domains are hence not entirely equivalent. In addition to spectral signatures, we observe clear (albeit very small) deviations in the lattice displacements, as measured from STM topographs (see SI for details).

\begin{figure}[!h]
    \centering
    \includegraphics[width =1 \columnwidth]{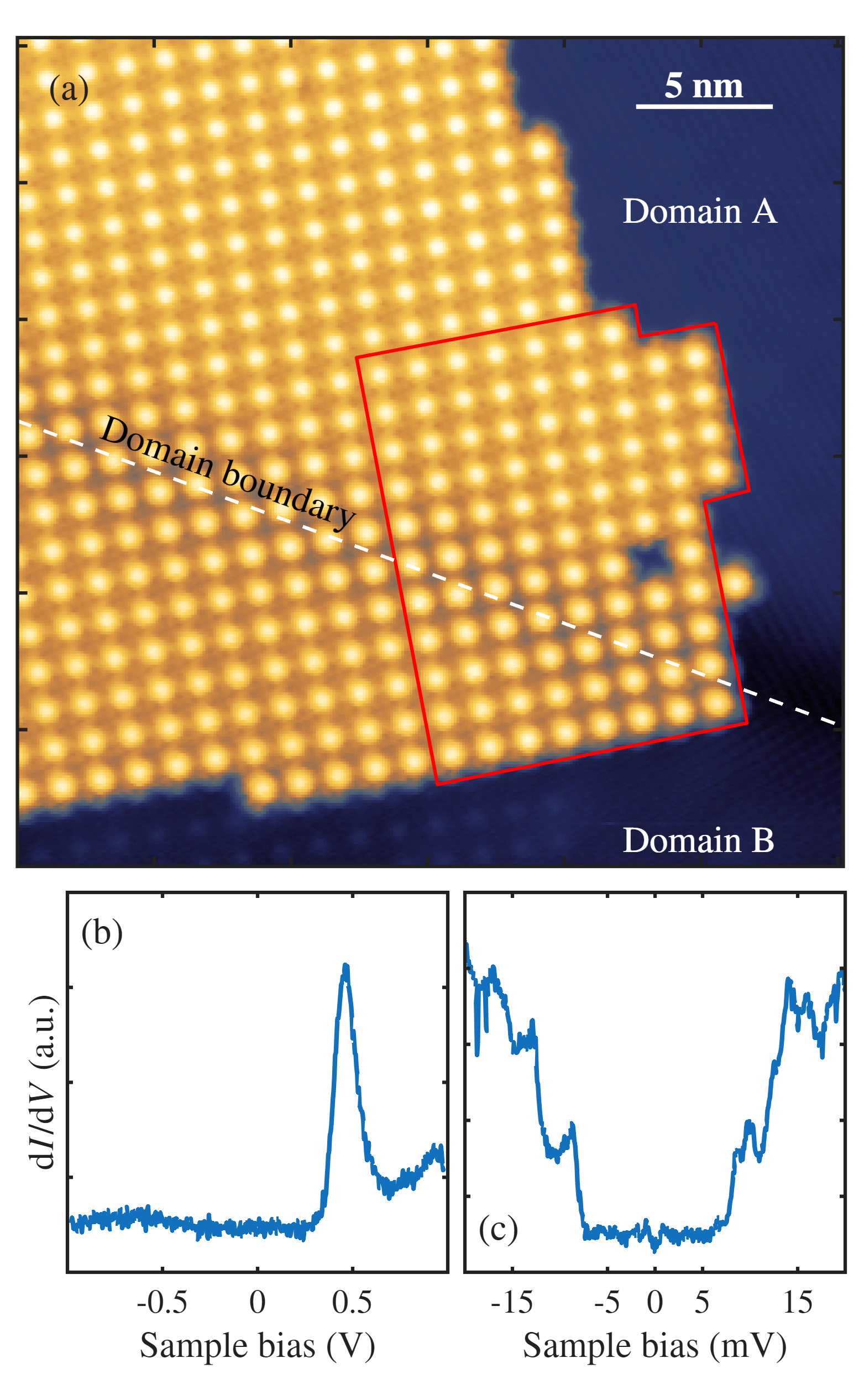}
    \caption{(a) STM image of an FePc island on BL SnTe. The molecules self-assemble into islands with a square lattice geometry (1\,V, 50\,pA). (b) Representative conductance spectrum acquired above the FePc centre ion showing the LUMO resonance with a maximum at ca. 460 mV. The LUMO maximum can be used to track the SnTe grain boundaries beneath the FePc island. (c) Representative conductance spectrum acquired above the centre ion of an FePc molecule showing the symmetric step-like increases of inelastic spin-flip excitations, encoding the spin properties of the molecules.}
    \label{fig:Figure2}
\end{figure}

We will now show that even these small changes in the SnTe lattice parameters result in clear and measurable changes to the magnetic properties of FePc. Like many metal phthalocyanines, FePc contains several unpaired electrons in its outer shell. The spin properties are generally retained even when adsorbed on surfaces \cite{hu2008electronic}. The excitation spectrum of the resulting spin system is experimentally accessible through inelastic tunnelling spectroscopy (IETS) \cite{Ternes2015}. It is also highly sensitive to the local environment through MCA. The ferroelectric domains of SnTe should thus induce clear changes in the excitation spectrum of FePc.

We demonstrate these changes by studying an  island of FePc molecules adsorbed over a domain boundary of SnTe, shown in \figref{fig:Figure2}a. The domain boundary is clearly visible as a depression in the SnTe layer in the lower right hand side of the image. A slight depression in the FePc island, aligned with the grain boundary, indicates that it continues beneath the molecular island. We denote the SnTe domain in the upper part of the image as domain A and the one in the lower part of the image as domain B. The approximate location of the domain boundary is indicated by a dashed line. 

\begin{figure}
    \centering
    \includegraphics[width=1.00\columnwidth]{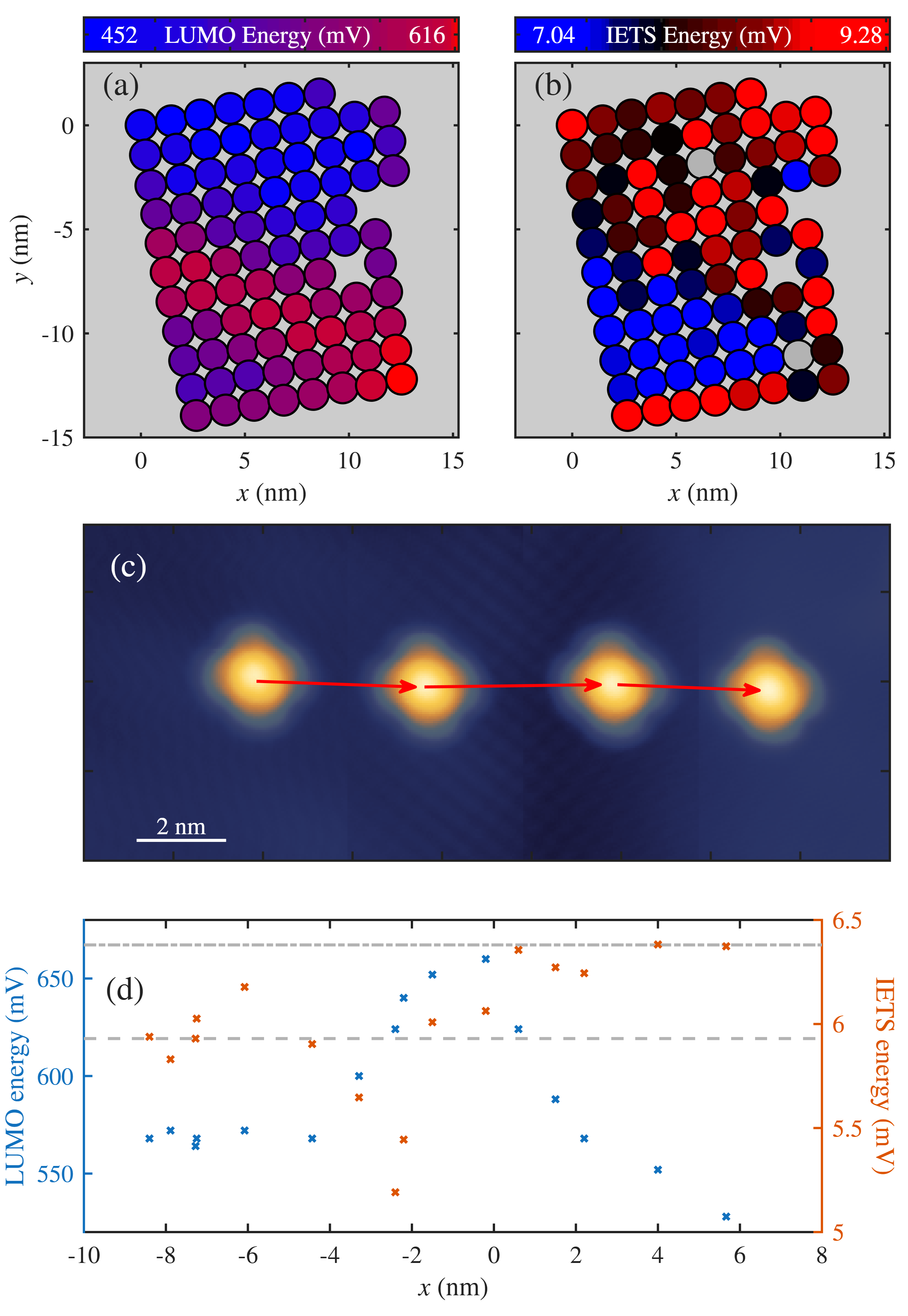}
    \caption{(a) Spatial map of the LUMO maxima of the molecules inside the red contour in \figref{fig:Figure2}a. Each circle represents one FePc molecule. The location of the grain boundary is clearly visible as an increase in the LUMO energy. (b) Spatial map of the lowest spin excitation energy as extracted from IETS spectra of the molecules inside the red contour in \figref{fig:Figure2}a. Each circle represents one FePc molecule. The domain contrast of SnTe is clearly represented in the spin excitation energies. Grey circles represent outliers with abnormally low spin excitation energies. (c) Composite STM image showing some steps of the molecular manipulation of an FePc molecule across the domain boundary next to the FePc island in \figref{fig:Figure2}a. Red arrows indicate the the manipulation sequence. Not all steps are shown. (d) LUMO maxima (blue) and lowest spin ecitation energies (orange) extracted at each point of the manipulation sequence. The origin of the $x$-axis is centered on the domain wall. The patterns observed in panels (a) and (b) for the FePc island also hold for single molecules.}
    \label{fig:Figure3}
\end{figure}

In order to correlate spin properties with assignment to a particular ferroelectric domain, it is first necessary to confirm the presence of the domain boundary underneath the FePc island and accurately track its trajectory. We do so using the lowest unoccupied molecular orbital (LUMO) of the FePc molecules as a marker, which appears as a pronounced peak in the range of ca. 450-600\,mV in conductance spectra acquired on the FePc molecules, as can be see in \figref{fig:Figure3}b. This variation in energy of the LUMO resonance is a response to the band bending at the domain boundary. We can thus track the domain boundary by mapping the location of the LUMO maximum.

We do so for the subset of molecules inside the red enclosure in \figref{fig:Figure2}a, measuring a conductance spectrum and extracting the LUMO maximum for each to produce a two-dimensional map of the LUMO energy. The result is shown in \figref{fig:Figure3}a, with each circle representing an FePc molecule and the LUMO energy represented by the colour scale. The modulation of the LUMO energy emulates that of the SnTe valence band described above: The LUMO shifts to higher energies above the domain boundary and settles to slightly different baselines in domains A and B. In addition, the LUMO is shifted to slightly higher energies at the edges of the island when compared with the bulk of the domains.
 
We now turn to the spin excitations, which manifest in the conductance spectra as a series of steps at characteristic energies symmetric with respect to zero bias \cite{heinrich2004single, hirjibehedin2006spin}. A representative example of the IETS spectrum of FePc on SnTe is shown in \figref{fig:Figure2}c. While the spin ground state of FePc/SnTe is either S=1 or S=2 based on DFT calculations (see SI for details), the zero field the zero field splitting and presence of  at least two clearly expressed steps shows that the molecule must carry a spin in excess of 1/2 \cite{Ternes2015}.

We focus our investigation on the lowest energy IETS step, using it as a marker to resolve modifications of the FePc energy diagram. As before, we acquire conductance spectra above the center of each of the molecules in the red enclosure in \figref{fig:Figure3}a, this time extracting the energy at which we observe the first spin transition. \figref{fig:Figure3}b shows the resulting map. The underlying domain structure of SnTe is readily apparent in the spin excitation map. The domain boundary appears as a gradual transition in the excitation energy and coincides with the independently determined boundary in the map of LUMO energies in \figref{fig:Figure3}a.

 It is clear from \figref{fig:Figure3}b that other factors also influence the energy of the first spin transition. For example, molecules at the edge of the island show a markedly different spin excitation energy, as do those next to the vacancy in the center right of the island, suggesting that inter-molecular coupling may also play a role in this system \cite{drost2023real, koch2025hamiltonian}.

 In order to rule out this inter-molecular interaction, or other factors independent of the SnTe, as a source of the observed pattern, we conducted a test experiment in which we detached a single molecule from the FePc island and manipulated it across the adjacent domain boundary, acquiring conductance spectra at each step (see \figref{fig:Figure3}c and SI). We repeat the same analysis as for the FePc island, extracting the center of the LUMO resonance and the first spin excitation energy at each step, thus tracing the electronic and magnetic properties of FePc as it migrates across the domain boundary. The results, shown in \figref{fig:Figure3}d, reproduce the behavior already observed for the FePc island.

The clear correlation of the spin excitation energy with the domain structure of SnTe, along with the fact that the same behavior persists in individual molecules, is a powerful indicator that FePc indeed responds to the ferroelectric properties of SnTe. We can trace this effect to small lattice distortions, most likely arising from subtle differences in stacking with the underlying HOPG substrate. These variations in lattice constant are also directly observable in atomically resolved STM images (see SI for details). We can thus assign the observed variations in spin excitation energy to the interplay of FePc with the ferroelectric properties of SnTe.

Our measurements demonstrate a structural mechanism to couple electric and magnetic degrees of freedom at the nanoscale. Our experiments give a proof of concept for designer multiferroics that leverage changes in the magneto-crystalline anisotropy. 
Two-dimensional multiferroics, along with their desirable properties, remain elusive. Rational material design in two-dimensional heterostructures has proven a powerful tool for creating relevant solid state systems, such as topological insulators or heavy fermion phases. The same concepts may be applied to generate designer multiferroics from tailored interactions of simpler materials. Our experiments show a plausible path towards a new class of engineered two-dimensional heterogeneous multiferroics by leveraging MCA to influence target spin systems.

\section*{Methods}

We grew SnTe samples by molecular beam epitaxy (MBE) in an ultra-high vacuum (UHV) chamber, choosing highly-oriented pyrolitic graphite (HOPG) as a substrate. The substrates were cleaved in UHV and annealed at 572\,K until the pressure in the UHV chamber stabilized. We deposited SnTe by direct sublimation of SnTe powder from a resistively heated crucible held at 822\,K onto the HOPG substrate, held at 482\,K. The deposition time was 3600\,s. We obtained large and clean islands of predominantly monolayer SnTe, with some bilayer areas, as was reported previously \cite{Chang2016, Amini2023}. After inspecting the samples for cleanliness, we deposited FePc at room temperature and annealed the sample at 472\,K for 600\,s to obtain well-ordered islands of FePc molecules. The focus of this paper are FePc molecules adsorbed on BL SnTe, though we think our findings to also be applicable to other thicknesses. All measurements were performed in a Unisoku USM-1300 STM operating at 350\,mK base temperature.

\section*{Data availability}

All data supporting the findings are available from the corresponding authors upon reasonable request.

\section*{Acknowledgements}

This research made use of the Aalto Nanomicroscopy Center (Aalto NMC) facilities and was supported by the European Research Council (ERC-2021-StG TITAN No.~101039500, ERC-2023-AdG GETREAL No.~101142364, ERC-2024-CoG ULTRATWISTROICS  No.~101170477), the Research Council of Finland (Academy Research Fellow No.~331342, No.~336243, No.~338478, No.~346654, No.~347266, No. ~369367, and No.~353839, the Finnish Quantum Flagship project No.~358877), InstituteQ and the computational resources provided by the Aalto Science-IT project. L.Y. acknowledges support from the Jiangsu Specially-Appointed Professors Program, Suzhou Key Laboratory of Surface and Interface Intelligent Matter (Grant SZS2022011), and Gusu Innovation and Entrepreneurship Talent Program - Major Innovation Team (ZXD2023002).

\section*{Competing interests}
The authors declare no competing interests.

\newpage
\phantom{Hidden text to force proper newpage}
\newpage

\setcounter{figure}{0}
\setcounter{equation}{0}
\newcommand{\suppcite}[1]{\cite[S\hspace{-1.8mm}][]{#1}}
\renewcommand\thefigure{S\arabic{figure}}
\renewcommand\theequation{S\arabic{equation}}

\makeatletter
	\renewcommand*{\@biblabel}[1]{[S#1]}
\makeatother

\begin{center}
\onecolumngrid
\section{Supporting Information for 'Atomic-scale probe of molecular magneto-electric coupling'}
\vspace{1cm}
\end{center}

\section*{Direct Measurements of \element{Sn}\element{Te} Lattice Distortion}

The minute distortions of the SnTe lattice are directly visible in atomically resolved STM images. To substantiate our hypothesis of varying magneto-crystalline anisotropy, we took several such images of SnTe domain boundaries to measure the SnTe lattice constant in various adjacent domains. The lattice distortion is directly evident in the data. We use the ratio of the shorter to longer lattice parameter to quantify the lattice distortion. This ratio changes from one domain to the next, as expected. Three instances of this type of measurement are shown in Figures S1 through S3, with extracted line profiles and the ratio of lattice parameters to demonstrate the lattice distortion.

\begin{figure}[!h]
    \centering
    \includegraphics[width=.8\linewidth]{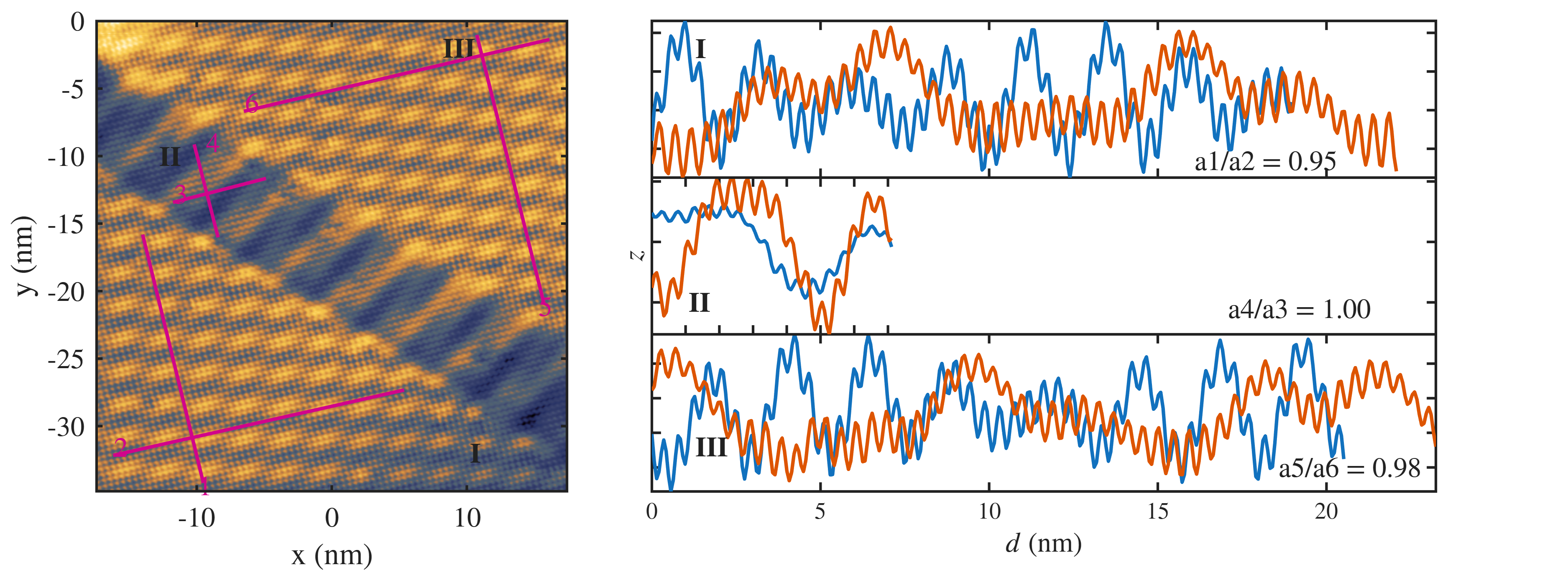}
    \caption{Atomically resolved STM image of several SnTe domains separated by a grain boundary. The lines mark the locations of the extracted line profiles on the right hand side.}
    \label{fig:FigureS1}
\end{figure}

\begin{figure}[!h]
    \centering
    \includegraphics[width=.8\linewidth]{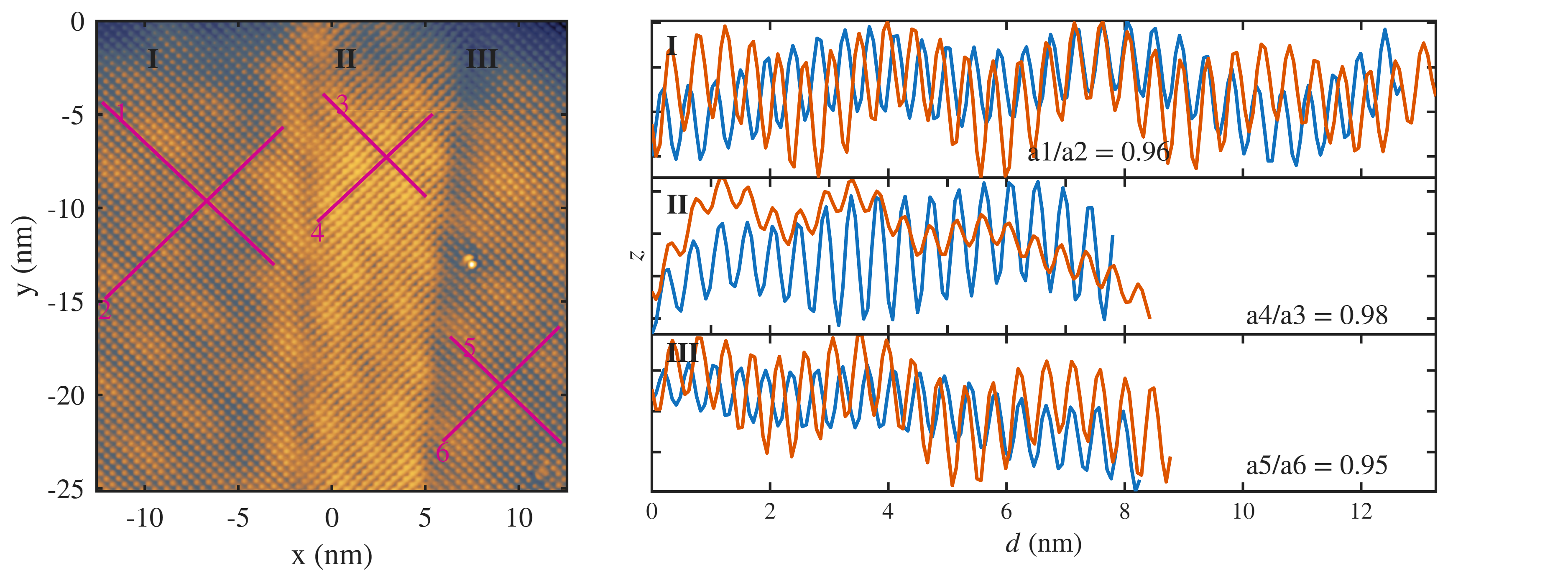}
    \caption{Atomically resolved STM image of several SnTe domains separated by a grain boundary. The lines mark the locations of the extracted line profiles on the right hand side.}
    \label{fig:FigureS2}
\end{figure}

\begin{figure}[!h]
    \centering
    \includegraphics[width=.8\linewidth]{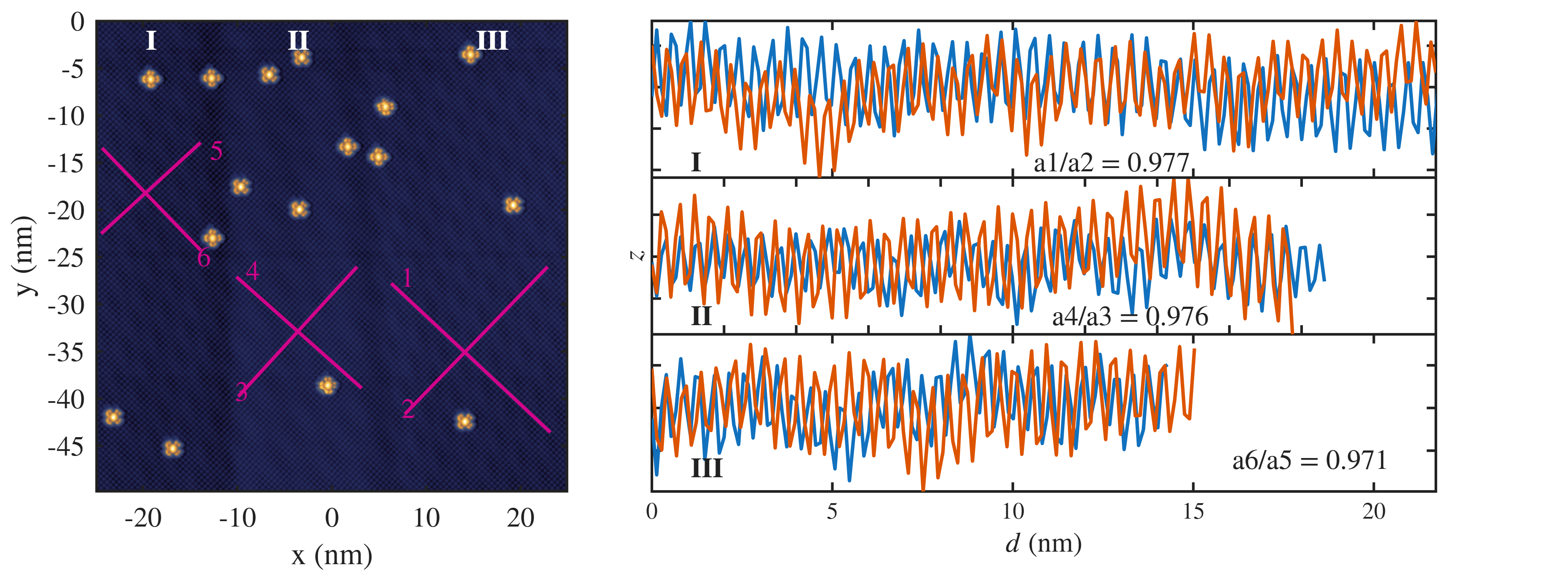}
    \caption{Atomically resolved STM image of several SnTe domains separated by a grain boundary. The lines mark the locations of the extracted line profiles on the right hand side.}
    \label{fig:FigureS3}
\end{figure}

\newpage
\section{Manipulation of \element{Fe}Pc Molecules}

We developed the following procedure for manipulation individual FePc molecules on SnTe:

\begin{enumerate}
    \item Position the tip above the centre of the target molecule
    \item Decrease the bias voltage to 10\,mV
    \item Increase the setpoint current to 1\,nA
    \item Re-position the tip to the target location with the feedbac engaged
    \item Restore the original bias and current values to release the molecule
\end{enumerate}

\begin{figure}[!h]
    \centering
    \includegraphics[width=.8\linewidth]{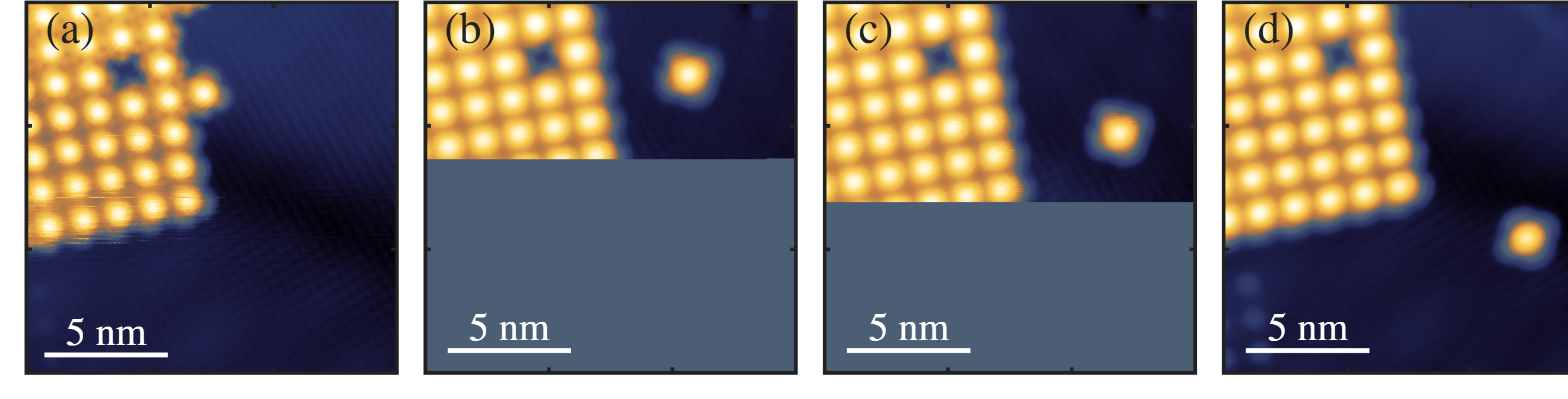}
    \caption{Representative manipulation sequence showing the detachment of a FePc molecule from a larger island and displacement towards the grain boundary.}
    \label{fig:figureS4}
\end{figure}

\begin{figure}[!h]
    \centering
    \includegraphics[width=.8\linewidth]{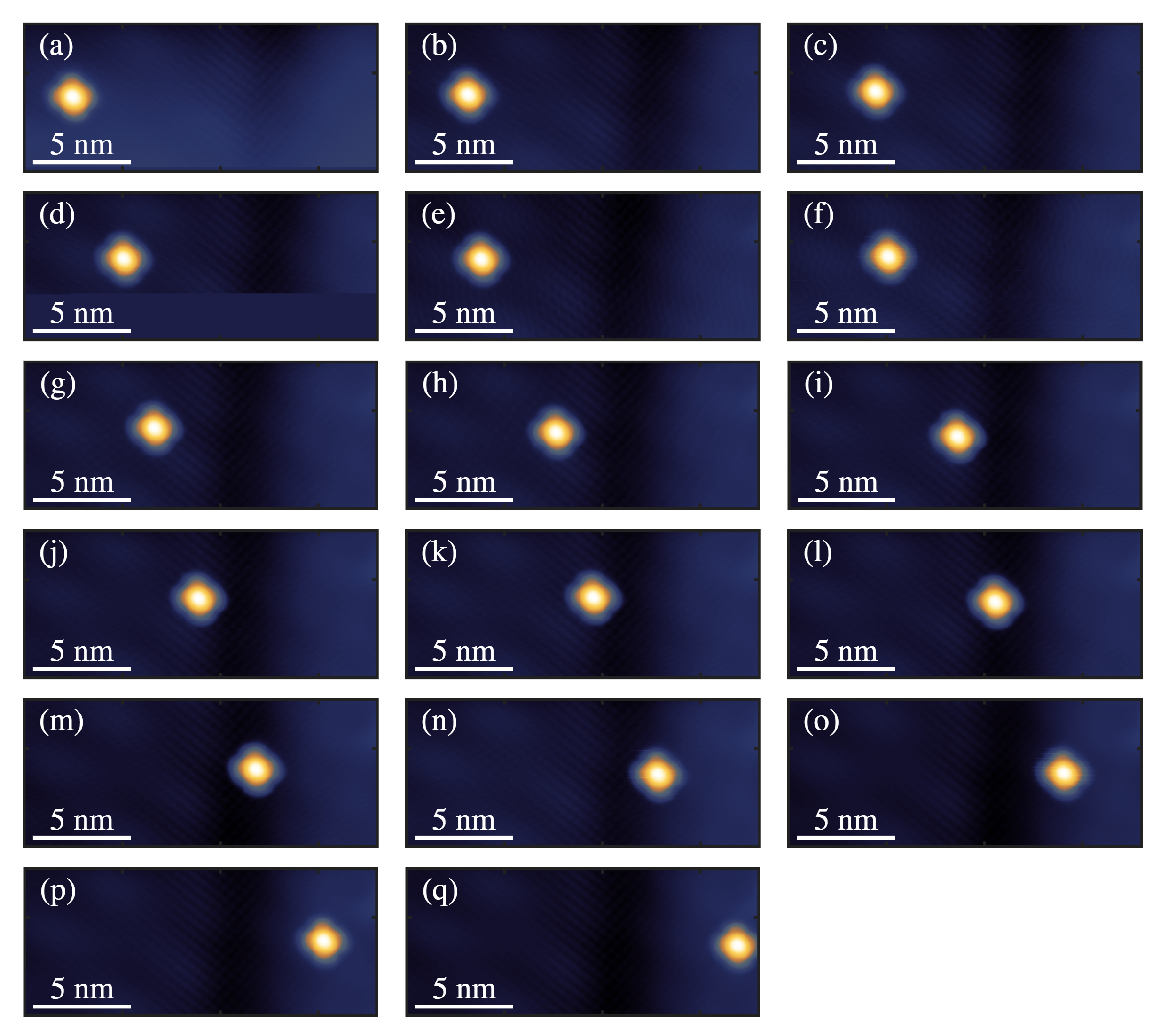}
    \caption{Representative manipulation sequence showing the passage of the FePc molecule across the SnTe grain boundary.}
    \label{fig:figureS5}
\end{figure}

This approach will move individual molecules with high fidelity. Molecules may come to rest at different, but discrete, angles with respect to the high-symmetry axes of the crystal when released. This appears to  be a stochastic process. The original rotation can be restored by performing minimal manipulations until finding the desired result. Some frames showing the detachment of the molecule from the island and the manipulation across the grain boundary are shown in \figref{fig:figureS4} and \figref{fig:figureS5}.

\section*{DFT calculations}

\begin{figure}[!h]
    \centering
    \includegraphics[width=1.\linewidth]{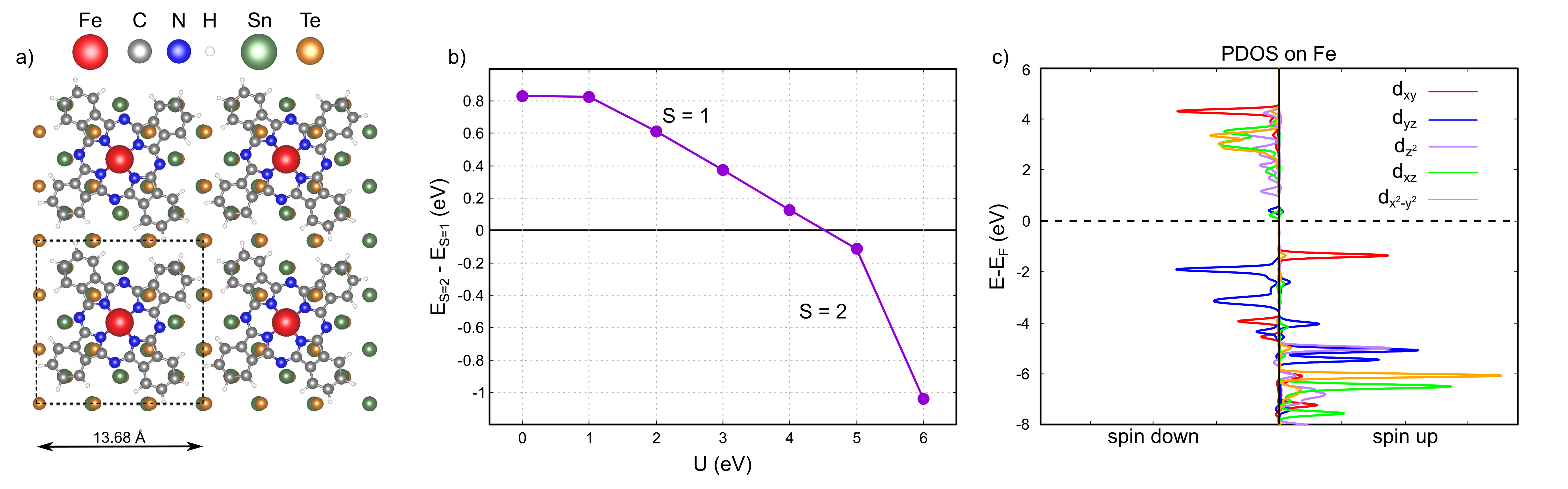}
    \caption{a) Top view of the optimized geometry of the FePc/SnTe system. The FePc molecule is positioned to achieve an intermolecular Fe–Fe distance of $\sim$ 14 Å. b) Evolution of the spin ground state energy of Fe as a function of the Hubbard $U$ parameter, showing a transition from a triplet ($S = 1$) to a quintuplet ($S = 2$) state for $U > 4$. c) Orbital-projected density of states (PDOS) for the Fe atom in the $S = 2$ ground state with $U = 6$ eV, showing the contributions of the 3d orbitals.}
    \label{fig:pdos}
\end{figure}

We performed spin-polarized density functional theory (DFT) calculations employing the DFT+U approach, with a Hubbard U correction applied to the Fe 3d orbitals, to accurately describe the strongly correlated electronic structure of the FePc molecule adsorbed on a SnTe monolayer. The top view of the FePc+SnTe geometry used in the DFT is shown in Figure \ref{fig:pdos}a, where we positioned the molecule in such a way that the intermolecular Fe-Fe distance is about 14 Å. Figure \ref{fig:pdos}b shows that the spin ground state changes drastically as the value of $U$ increases within the DFT+U approximation. For $U$ larger than 4 eV and in the absence of spin-orbit coupling, the spin configuration of the Fe atom in the FePc molecule changes from a triplet $S = 1$ to the quintuplet $S = 2$, where for $U = 6$ eV the $S = 2$ spin state is more energetically favorable by more than 1 eV. Figure \ref{fig:pdos}c shows the orbital character of the Fe atom in the S = 2 state with $U = 6$ eV in a large energy range between the Fermi level (-8 to 6 eV).

\subsection*{Computational details}

DFT calculations were performed using the periodic plane-wave-basis VASP code \cite{kresse1996,kresse1996_2}, with PAW pseudopotentials. Atomic positions and lattice parameters were obtained by fully relaxing all structures using the spin-polarized Perdew-Burke-Ernzehof (PBE) functional\cite{perdew1996} including dispersion correction with Becke-Johnson damping function (DFT-D3) \cite{grime2011}. An energy cut-off of 400 eV was used to expand the wavefunctions, and a systematic k-point convergence was checked for all structures, with sampling chosen according to each system size. The convergence criterion of self-consistent field computation was set to $10^{-5}$ eV, and the threshold for the largest force acting on the atoms was set to less than 0.01 eV Å$^{-1}$. A vacuum layer of 12 Å was added to avoid mirror interactions between periodic images. The spin polarization was considered in all calculations, and tested by varying the U parameter within the DFT+U approach introduced by Dudarev \textit{et al} \cite{dudarev1998}. All systems were relaxed again for different $U$ values.

\end{document}